\documentclass{iau} 
\usepackage{epsfig}
\usepackage{graphicx}

\title[Magnetic O-star H$\alpha$ Mass-Loss Rates] 
{First Empirical Constraints on the Low H$\alpha$ Mass-Loss Rates of Magnetic O-Stars}

\author[F.~A.~Driessen, J.~O.~Sundqvist \& G.~A.~Wade]   
{Florian A.~Driessen$^1$, Jon O.~Sundqvist$^1$ \and Gregg A.~Wade$^2$}

\affiliation{$^1$Institute of Astronomy, KU Leuven, \\ Celestijnenlaan 200D box 2401,
BE-3001, Leuven, Belgium \\ email: {\tt florian.driessen@kuleuven.be} \\[\affilskip]
$^2$Dept. of Physics \& Space Science, Royal Military College of Canada, \\ PO Box 17000, Station Forces, Kingston, Ontario, Canada}

\pubyear{2018}
\volume{346}  
\jname{High-mass X-ray binaries:
illuminating the passage from massive binaries
to merging compact objects}
\editors{L.~Oskinova, E.~Bozzo, D.~Gies, \& D.~Holz, eds.}
\begin{document}

\maketitle

\begin{abstract}

A small subset of Galactic O-stars possess surface magnetic fields that alter the outflowing stellar wind by magnetically confining it. Key to the magnetic confinement is that it induces rotational modulation of spectral lines over the full EM domain; this allows us to infer basic quantities, e.g., mass-loss rate and magnetic geometry. Here, we present an empirical study of the H$\alpha$ line in Galactic magnetic O-stars to constrain the mass fed from the stellar base into the magnetosphere, using realistic multi-dimensional magnetized wind models, and compare with theoretical predictions. Our results suggest that it may be reasonable to use mass-feeding rates from non-magnetic wind theory if the \textit{absolute} mass-loss rate is scaled down according to the amount of wind material falling back upon the stellar surface. This provides then some empirical support to the proposal that such magnetic O-stars might evolve into heavy stellar-mass black holes (\cite[Petit et al. 2017]{Petit_etal17}).

\end{abstract}

\firstsection 
\section{Introduction}

Hot, luminous, massive OB stars are known to have high-speed, radiation driven winds (\cite[Castor et al. 1975]{Castor_etal75}). Spectropolarimetric surveys over the past decade have shown that about 7\% of massive stars harvest a strong, often dipolar, surface magnetic field (\cite[Wade et al. 2016]{Wade_etal16}). The wind-magnetic field interaction around these stars leads to magnetic confinement of the wind and the formation of a circumstellar magnetosphere (\cite[ud-Doula \& Owocki 2002]{Asif02}). For the slowly rotating magnetic O-stars studied here, trapped material falls back onto the star on a dynamical timescale and the formation of a Dynamical Magnetosphere (DM) occurs (\cite[Sundqvist et al. 2012]{Sundqvist_etal12}; \cite[Petit et al. 2013]{Petit_etal13}). The DM structure is the origin of a range of spectral line diagnostics which we study here by means of high-quality, high-resolution spectra acquired by the Magnetism in Massive Stars survey (MiMeS; \cite[Wade et al. 2016]{Wade_etal16}).

The predicted fall-back of wind material upon the star leads naturally to the question if these magnetic stars lose less mass than their non-magnetic counterparts, and so might evolve into Galactic heavy mass black holes when ending their lives (\cite[Petit et al. 2017]{Petit_etal17}). However, so far models studying this have assumed that the wind mass fed into the magnetosphere was unaltered by the presence of the magnetic field. Hence, the mass launched from the magnetic star's base (the feeding rate) is assumed to be equal to that of a non-magnetic star (\cite[Vink et al. 2000]{Vink_etal00}). To examine this assumption empirically, here we perform a systematic study of H$\alpha$ diagnostics of a small sample of confirmed Galactic magnetic O-stars. 

\section{H$\alpha$ line-formation and mass-loss rates}

To compute theoretical H$\alpha$ line-profiles we follow the procedure of \cite[Sundqvist et al. (2012)]{Sundqvist_etal12} to solve the formal solution of radiative transfer in 3D cylindrical space for an observer viewing under angle $\alpha$ w.r.t.~the magnetic axis. We complement the 3D radiative transfer with a description of the DM as provided by the Analytical Dynamical Magnetosphere (ADM) formalism (\cite[Owocki et al. 2016]{Owocki_etal16}) to describe the velocity and density inside the DM. 

Due to magnetic confinement the mass that can effectively escape the star $\dot{M}_B$ is much smaller than the mass that gets launched from the base of the star $\dot{M}_{B=0}$. The rates are related via $\dot{M}_B = f_B \dot{M}_{B=0}$, with $f_B$ essentially the fraction the magnetosphere covers (\cite[ud-Doula et al. 2008]{Asif08}; their Eq.~23).

\section{Sample study and empirical H$\alpha$ mass-loss rate constraints}

\begin{figure}
\centering
\includegraphics[scale=0.25]{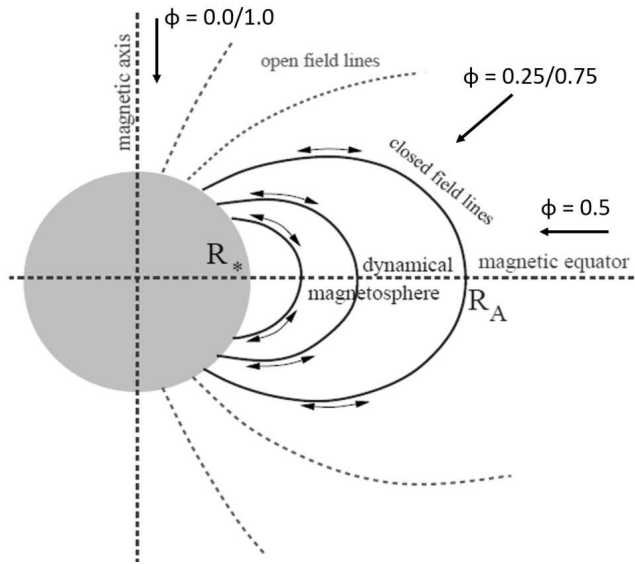}
\caption{Cartoon of a massive star with oblique dipolar magnetic field. Because of magnetic obliquity the stellar rotation results in a change of projected surface area of the magnetosphere, i.e., the angle between observer and magnetic field axis changes in time. This leads to modulated line-profile emission with rotation phase $\phi$.}
\label{fig:rotmod}
\end{figure}

The rotationally phase-modulated emission of the H$\alpha$ line (Figure \ref{fig:rotmod}) can be used to infer and constrain both the magnetic geometry $(i,\beta)$, setting the shape of the rotationally modulated emission, and the mass-feeding rate $\dot{M}_{B=0}$ that primarily affects the absolute amount of H$\alpha$ emission.

\begin{figure}
\centering
\includegraphics[scale=0.45]{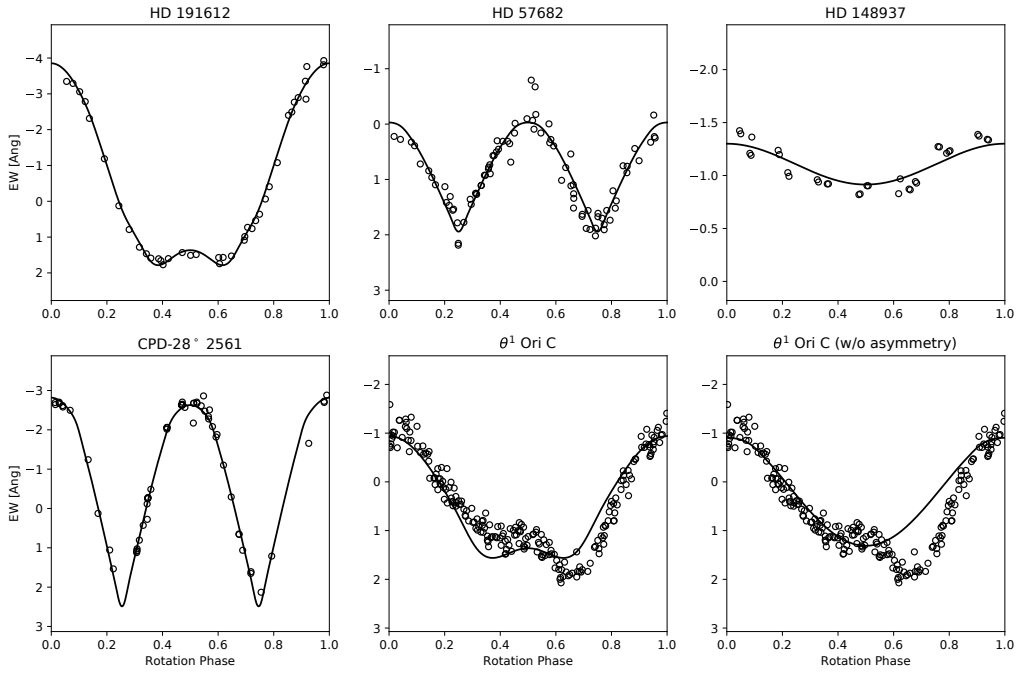}
\caption{Fits of phased H$\alpha$ EW variations of the investigated sample of confirmed Galactic magnetic O-stars by the MiMeS survey. Open black dots show observations and the black solid line is the ADM best-fit theoretical profile.}
\label{fig:fits}
\end{figure}

Figure \ref{fig:fits} displays the fitted phased H$\alpha$ equivalent widths (EW) of our sample stars, showing overall good agreement between the ADM model and observations. We note that $\theta^1$ Ori C exhibits an asymmetry in its lightcurve; the origin of this feature is unknown, but might be due to star spots or non-dipolar field contributions (\cite[ud-Doula et al. 2013]{Asif13}), both which cannot currently be modelled with the ADM. Because we are primarily interested in constraining mass-feeding rates we leave the asymmetry out (effectively by fitting only the region $0\leq \phi\leq 0.5$) as the absolute amount of emission is only marginally affected by this (Figure \ref{fig:fits}, panel 5 \& 6). 

\newpage 

The ADM is a steady-state model that predicts a smooth wind outflow. However, more realistically a transient infall of matter occurs onto the star which leads to statistical density variations. These are here modelled by a clumping factor $f_{cl}=\langle \rho^2 \rangle / \langle \rho \rangle ^2$, which scales down the mass-loss rate by $\sqrt{f_{cl}}$ appropriate for $\rho^2$-diagnostics like H$\alpha$ (e.g., \cite[Puls et al. 2008]{Jo2008}). 

\begin{figure}
\centering
\includegraphics[scale=0.6]{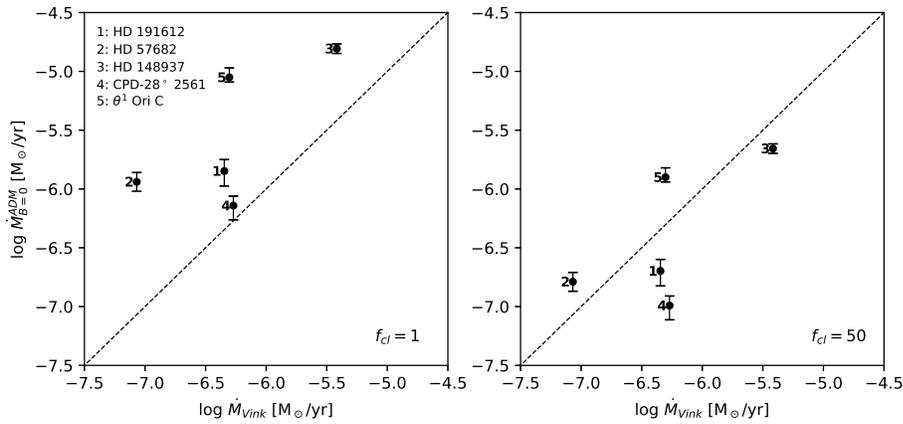}
\caption{H$\alpha$ ADM non-magnetic mass-feeding rates as a function of the non-magnetic mass-feeding rates predicted by \cite[Vink et al. (2000)]{Vink_etal00}. The black dashed line is a one-to-one correspondence. Error bars are 3$\sigma$ confidences.}
\label{fig:mdots}
\end{figure}

Time-dependent magnetohydrodynamic simulations predict $f_{cl}\approx 50$ for a magnetic O-star (\cite[Owocki et al. 2016]{Owocki_etal16}). We apply the above correction to our constrained $\dot{M}_{B=0}$ (from Figure \ref{fig:fits}) and compare with the \cite[Vink et al. (2000)]{Vink_etal00} prescription (Figure \ref{fig:mdots}). The latter are computed according to the stellar parameter compilation of \cite[Petit et al. (2013)]{Petit_etal13}. Though the scatter is large, the indication is that such a clumping factor indeed gives a rather small overall off-set, thereby lending some support to previous studies that have used the Vink et al.~prescription for $\dot{M}_{B=0}$ in their studies of magnetic O-stars (e.g., \cite[Petit et al. 2017]{Petit_etal17}). We emphasize, however, that the absolute mass-loss rate is much lower than the mass-feeding rate; e.g., for HD 191612 one gets $\dot{M}_B = f_B \dot{M}_{B=0} \approx 10^{-8}$ $M_\odot/$yr when using an Alfv\'en radius $R_A=3.5R_\star$ (\cite[Owocki et al. 2016]{Owocki_etal16}). 

\newpage

\section{Conclusions}

Although further observations and modelling efforts are certainly needed to draw more firm conclusions, our pilot-study here nevertheless provides some first empirical support that non-magnetic massive star mass-feeding rates can also be used in studies of magnetic massive stars. This result may then also lend some support to similar studies in magnetic massive star evolution (e.g., \cite[Petit et al. 2017]{Petit_etal17}), provided that the \textit{absolute} mass-loss rate $\dot{M}_B$ is scaled down accordingly (see Section 2). 

Currently additional investigations are being performed to better assess the uniqueness of our best-fit solutions and the influence of NLTE effects; results from this will be presented in an upcoming paper (Driessen et al., in prep.).

\end{document}